# State Feedback Control for Adjusting the Dynamic Behavior of a Piezo-actuated Bimorph AFM Probe

Bilal Orun[1], Serkan Necipoglu[2], Cagatay Basdogan[1*] and Levent Guvenc[2]

[1]College of Engineering, Koc University, Istanbul 34450, Turkey

[2]Department of Mechanical Engineering, Istanbul Technical University, Istanbul 34437, Turkey

**Abstract:** We adjust the transient dynamics of a piezo-actuated bimorph Atomic Force Microscopy (AFM) probe using a state feedback controller. This approach enables us to adjust the quality factor and the resonance frequency of the probe *simultaneously*. First, we first investigate the effect of feedback gains on dynamic response of the probe and then show that the time constant of the probe can be reduced by reducing its quality factor and/or increasing its resonance frequency to reduce the scan error in tapping mode AFM.

## I. INTRODUCTION

The scan performance in a typical AFM system is related to the dynamical performance of the cantilever probe and the scanner. Given a piezo scanner with a high mechanical bandwidth and a robust controller that works with it, the scanning bandwidth and the image resolution are mainly governed by the cantilever dynamics. Fantner et al.[1] emphasized that the maximum scan speed is determined by the spring constant, the effective mass of the cantilever, the damping of the cantilever in the surrounding medium, and the sample stiffness. Albrecht et al.[2] reported that the response of a cantilever probe may be expressed in terms of a time constant $\tau = 2Q/\omega_n$, where $Q$ is the quality factor and $\omega_n$ is the n$^{th}$ resonance frequency of the probe. Hence, it is desired to have low values of $Q$ and high values of resonance frequency for faster response. Increasing the Q factor of a cantilever probe limits its bandwidth since the time constant of the probe is inversely proportional to its "sensing" bandwidth. Albrecht et al.[2] showed that for a cantilever probe in vacuum having high $Q$ factor ($Q = 50000$) and a typical resonant frequency of 50 kHz, its maximum available bandwidth is only 0.5 Hz, which is not usable for most dynamic AFM applications. Mertz et al.[3] introduced a method that allows the active modification of the cantilever's damping by the controlled increase or decrease of the apparent (effective) $Q$ factor of the system, also known as the "$Q$ control" today. In $Q$ control, the displacement of the probe is first measured using a sensor, then the signal is phase shifted in the time domain to obtain the corresponding velocity, finally the velocity signal is multiplied by a gain factor, $G$, and then added (subtracted) to (from) the actuation signal to change the effective damping (also the effective $Q$ factor) of the cantilever probe. This can be achieved using an additional electronic circuit, including a phase shifter and a gain amplifier in the feedback loop. Sulchek et al.[4] showed that the sensing bandwidth of a scanning probe and its scan speed can be significantly improved by using a piezoelectric probe (instead of using a piezo-tube to actuate the probe in the z-axis) and then by actively lowering the $Q$ factor of the probe when scanning nano surfaces in air. Active damping allows the amplitude of the oscillating cantilever to respond faster to topographical changes at the expense of lower force sensitivity of the cantilever. Rodriguez and Garcia[5] developed an analytical model of a cantilever probe in the form of a mass-spring-damper system and investigated its dynamics under $Q$ control in tapping mode AFM. They emphasized the importance of the transients in the cantilever's response and showed that the active response of a cantilever probe can be increased or decreased depending on the phase shift of the self-excitation. The maximum change in Q occurs when the value of phase shift is at $\phi = \pm 90$ degrees. Moreover, they reported that $Q$ enhancement reduces the maximum force exerted by the tip on the sample surface. Chen et al.[6] changed the $Q$ factor of a cantilever probe using $Q$ control and showed that increased effective $Q$ promotes the attractive regime, improves imaging sensitivity, and results in less invasive imaging of soft biological samples. Numerical simulations performed by Kokavecz et al.[7] also support this argument. Ebeling et al.[8] compared imaging in liquid with and without the $Q$ control and observe that heights measured with active $Q$ control are reproducibly higher as compared to the ones observed without $Q$ enhancement. This effect is attributed to the reduction of tip-sample forces by $Q$ control. In fact, Jaggi et al.[9] experimentally determined that the average tip-sample forces are reduced by $Q$ enhancement. Despite these benefits, $Q$ enhancement increases the transient time and adversely affects the scan speed. Holscher and Schwarz[10,11] developed a mathematical model of a cantilever probe under $Q$ control and determine the theoretical limits of the gain factor used for adjusting the effective $Q$ factor of the probe in tapping mode AFM. They emphasized that adjusting the phase shift between the excitation and the cantilever oscillations to modify the $Q$ factor of the probe could be problematic in real experiments and show that the gain factor is limited by $1/Q$ when the phase shift is $\pm 90$ degrees and the effective $Q$ factor of the probe becomes $Q_{eff} = 1/(1/Q \pm G)$, where $G$ is the gain factor. When the same phase shift is increased to 0 or 180 degrees (corresponds to pure position signal), the native $Q$ factor of the probe does not change but the resonance frequency is shifted by $0.5(f_n G)$, where $f_n = \omega_n/2\pi$ is the resonance frequency of the probe.

Gunev et al.[12] suggested a new approach called adaptive $Q$ control (AQC), in which the $Q$ factor of a piezoelectric probe is adjusted on the fly during scanning. In standard $Q$ control, achieving higher scan speeds with reduced tapping forces is not possible since the effective $Q$ factor of the probe is set to a value that is lower or higher than its native one before scanning. In AQC, the controller modifies the $Q$ factor of the probe on the fly to avoid the error saturation problem which typically occurs in scanning steep downward steps. The results of the experiments performed with an AFM setup showed that the performance of AQC is superior to that of the standard $Q$ control. Varol et al.[13] performed numerical simulations in SIMULINK to investigate nano scanning in tapping mode AFM under $Q$ control. They focused on the simulation of the whole scan process rather than the simulation of cantilever dynamics and the force interactions between the probe tip and the surface alone, as in most of the earlier numerical studies. They discussed the trade-off in setting $Q$ factor of the probe in $Q$ control (i.e. low values of $Q$ cause an increase in tapping forces while higher ones limit the maximum achievable scan speed due to the slow response of the cantilever to the rapid changes in surface profile) and showed the differences in scan performance at different settings using the iso-error curves obtained from the numerical simulations.

Most of the earlier studies on controlling the dynamical response of a cantilever probe have focused on adjusting its $Q$ factor (i.e. damping). However, the stiffness and the mass of the probe also play a crucial role in the response since they directly affect the resonance frequency ($\omega_n = \sqrt{k/m}$, where $k$ and $m$ are the effective stiffness and mass of the probe respectively). Viani et al.[14] showed that the response time of a cantilever probe can be improved using smaller cantilevers. As a result, the mass of the cantilever is reduced, and the resonance frequency is increased, leading to a smaller time constant and faster response. However, using smaller cantilevers may cause difficulties while engaging sample surfaces and when scanning surfaces having large topographic variations. Varol et al.[13] showed that scanning with soft cantilevers having high effective $Q$ factor results in a better image quality especially when scanning soft samples since the force sensitivity of the cantilever increases. Moreover, the risk of damaging biological samples is reduced. However, the response time of a soft cantilever with high $Q$ factor is poor if the relation given by Albrecht et al.[2] for the time constant ( $\tau = 2Q/\omega_n$ ) is considered.

High or low values of $Q$ factor and resonance frequency may be desirable depending on the application. For example, if the purpose is to increase the scan speed in air, low $Q$ and high $\omega_n$ is preferred. On the other hand, if a biological sample is to be scanned in liquid, $Q$ factor must be enhanced for higher force sensitivity and the effective stiffness of the probe can be reduced to prevent damaging the sample. As is obvious from the above discussion, adjusting the effective $Q$ factor (by changing the damping) and the resonance frequency $\omega_n$ (by changing the stiffness) of a cantilever probe is highly beneficial. State feedback control is necessary in order to change the effective stiffness and damping of an AFM probe by altering the states of the probe simultaneously, The state space approach has been recently applied to modeling cantilever dynamics in AFM studies. Stark et al.[15] constructed a state space model of a cantilever probe to investigate its dynamic response. They integrated the nonlinear forces between the tip and sample surface into the model as output feedback. This approach enabled them to study the complex dynamics of different AFM modes through numerical simulations within one unified model. To capture the transient dynamics of a cantilever probe, Sebastian et al.[16] developed a state space model of the probe first and then estimated its velocity using a state observer. An observer is a computer implemented mathematical model that enables the estimation of a physical state that may not be measured directly. In typical AFM measurements, the probe position is available but the velocity of the probe is not measured. Knowing the full state of the probe is helpful to better understand the tip-sample interactions and also opens the door for the implementation of new control strategies in dynamic AFM as well. In a later study, Sahoo et al.[17] showed how the $Q$ factor of the probe can be altered using a velocity observer. They emphasized the trade-off between resolution and bandwidth in $Q$ control and show that observer-based $Q$ control method provides greater flexibility in managing this trade-off.

In this study, we present experimental methods for adjusting the effective $Q$ factor and the resonance frequency of a cantilever probe through state feedback (Fig. 1). To achieve this, we first measure the velocity of the oscillating probe using a Laser Doppler Vibrometer (LDV) and then obtain the position signal from the velocity signal via an analog integrator circuit. We finally multiply the velocity signal with a gain $G$, the position signal with a gain $H$, and then feed them back to the probe to change its effective damping and the stiffness, respectively. In order to make this change *simultaneously*, we utilize state feedback control. While changing the damping of the probe just affects its $Q$ factor, changing the stiffness not only affects the resonance frequency but also the $Q$ factor of the probe at the same time. Hence, state feedback approach is crucial for the calculation of the proper feedback gains. Using the calculated feedback gains, we perform scanning experiments to investigate the effect of state feedback on the image quality.

In addition to the scanning experiments, we further investigate the influence of state feedback on the dynamic response of the probe through numerical simulations performed in SIMULINK. For this purpose, we obtain a transfer function of the probe using the experimental data collected through frequency sweeping. We then investigate the effect of feedback gains on the scan speed, the tapping forces, and the image quality. We also show the trade-off between the scan speed and the tapping forces under state feedback control, which suggests that the feedback gains $H$ and $G$ must be selected carefully for optimum scan performance.

********* Insert Figure 1 here *************************************

## II. STATE FEEDBACK

Full state feedback using pole placement design, is a method employed in feedback control system theory to place the closed loop poles of a plant in predetermined locations in the complex plane. The location of the poles is related to the selection of the eigenvalues of the system, which determine the characteristics of the dynamical response of the system.

If we model the probe as a mass-spring-damper system vibrating under the influence of external forces, then its equation of motion can be written as:

$$m\ddot{z} + b\dot{z} + kz = F(t) \tag{1}$$

where, $m$, $b$, and $k$ represent the effective mass, damping, and the stiffness of the probe respectively and $F(t)$ represents the sinusoidal force applied to the probe tip for actuation. The transfer function between the force applied to the probe tip, $F(s)$, and the position of the probe tip, $z(s)$, can be written in the Laplace domain as:

$$\frac{z(s)}{F(s)} = \frac{1/m}{s^2 + (b/m)s + (k/m)} \tag{2}$$

Now, if we consider a linear relation between the voltage applied to the piezo layer of the probe and the corresponding force generated at the probe tip, $F(s) = B_0 V(s)$, where $B_0$ is a constant, then the transfer function between the voltage applied to the piezo layer, $V(s)$, and the tip position, $z(s)$, can be written as:

$$\frac{z(s)}{V(s)} = \frac{B_1}{s^2 + A_1 s + A_0} \tag{3}$$

where, $A_0 = \omega_n^2 = k/m$, $A_1 = \omega_n/Q = 2\zeta\omega_n = b/m$, and $B_1 = B_0/m$.

This transfer function can be represented by a state space equation as:

$$\dot{x} = Ax + BV$$
$$y = Cx \tag{4}$$

where, $x = [z \quad \dot{z}]^T$ is the state vector representing the position and velocity of the vibrating probe tip, $V$ is the control input (i.e. applied voltage), $y$ is the output vector, and the matrices $A$, $B$, and $C$ are defined as

$$A = \begin{bmatrix} 0 & 1 \\ -A_0 & -A_1 \end{bmatrix}, \quad B = \begin{bmatrix} 0 \\ B_1 \end{bmatrix}, \text{ and } C = [0 \quad 1] \tag{5}$$

The poles of the system are the roots of the characteristic equation given by $|sI - A| = 0$. Full state feedback is achieved by defining an input proportional to the state vector as $V = -Kx$, where $K = [H \quad G]$ is the gain vector. Now, the roots of the full state feedback system are given by the roots of the characteristic equation $|sI - (A - BK)| = 0$. By selecting proper gains for the vector $K$, we can place the poles of the closed loop system at the desired locations. If Ackermann's approach is utilized for the pole placement, the gain vector $K$ is calculated as

$$K = [0 \quad 1] P_C^{-1} q(A) \tag{6}$$

where, $P_c = [B \quad AB]$ is the controllability matrix and $q(A) = A^2 + A_1^{des} A + A_0^{des} I$ is the desired characteristic equation with $A$ in place of $s$. The rank of the matrix $P_c$ is 2 in our case, indicating a fully state controllable system. Hence, the vector $K$ for our system can be calculated as

$$K = [H \quad G] = \frac{1}{B_1} \left[ A_0^{des} - A_0 \quad A_1^{des} - A_1 \right] = \frac{m}{B_0} \left[ (\omega_n^{des 2} - \omega_n^2) \quad (\frac{\omega_n^{des}}{Q^{des}} - \frac{\omega_n}{Q}) \right]$$
$$= \frac{1}{B_0} \left[ (k^{des} - k) \quad (b^{des} - b) \right] \tag{7}$$

This result is not unexpected if the equation of motion of the probe, Eq. (1), is inspected carefully (note that $k^{des} = k \pm H$ and $b^{des} = b \pm G$ for position and velocity feedback).

## III. SETUP

To implement the proposed state feedback approach, we used a home made AFM operating in tapping mode. Details about the setup and the components are available in our earlier publication[12]. The major components of our AFM setup include a piezoelectric AFM probe which is brought close to a sample surface using an XYZ manual stage, a computer controlled XYZ nano stage for moving the sample surface with respect to the probe, and a LDV for measuring the vertical vibrations of the probe. An analog signal processing circuit consisting of (a) a root mean square (RMS) converter, (b) a variable phase shifter and (c) a voltage multiplier was built and integrated into our AFM setup in the past to adaptively modify the effective $Q$ factor of the probe on the fly during nano scanning[12]. Additionally, a simple analog integrator is built to obtain the position signal from the velocity signal measured by the LDV through integration and a voltage multiplier is added to the circuit for position feedback (see Fig. 2). The feedback gains $H$ and $G$ can be adjusted by means of the potentiometers shown in Fig. 2. The proper selection of gain values is crucial for attaining the desired values of the resonance frequency and the $Q$ factor simultaneously. In our system, the ratio of position gain to velocity gain, $H/G$, is very large since $k \gg b$ (see Eq. (7)). Hence, the gain factor $H$ is several orders of magnitude greater than $G$ and supplying gain $H$ directly through a voltage source is impossible. A large fraction of this gain is provided by the analog integrator circuit shown in Fig. 2. The output of the operational amplifier (LF351) in this figure is the integrated velocity signal (i.e. position), which is amplified by the integrator gain $1/(R3 \times C2)$. The large resistance $R4$ connected in parallel to the capacitor $C2$ in the integrator circuit is used for stability purposes only. Note that the resistance $R_{var}$ in Fig.2 is adjusted in advance such that the output coming from the phase shifter is always in phase with the true velocity signal[12].

### IV. ESTIMATING THE TRANSFER FUNCTION OF THE CANTILEVER PROBE

In order to estimate the feedback gains $H$ and $G$, which can be used in the electronic circuit shown in Fig. 2 to alter the resonance frequency and the $Q$ factor of the probe, first, a transfer function of the probe is developed based on the frequency sweep data and then the gains are calculated using the pole placement approach implemented in MATLAB. For this purpose, the probe is driven by a sinusoidal input voltage at different frequencies around the resonance frequency of 224 kHz (note that the first three resonance frequencies of the probe are far away from each other and there is no coupling between them[12]) and the amplitude of the probe velocity is measured experimentally using the LDV. We then developed an iterative curve fitting algorithm to identify the parameters of the transfer function given in Eq. (3). The parameters of interest in the transfer function are $A_0$, $A_1$, and $B_1$. Since the resonance frequency of the probe, $\omega_n$, can be determined directly via frequency sweep, $A_0$ is known in advance, and we only need to determine the parameters $A_1$ and $B_1$. The parameter $B_1$ affects the system gain only while $A_1$ affects both the gain and the damping ratio. We calculate $B_1$ and $A_1$ by curve fitting of the experimental amplitude curve. For this purpose, we iterate $A_1$ and $B_1$ until the estimated curve matches the experimental one with a small acceptable error (see Fig. 3). The error is defined as the difference between the areas under the estimated and experimental curves and the aim of the curve fit algorithm is to minimize this error down to an acceptable value through the iterations. Once the parameters of the transfer function are estimated, the state variable feedback gain vector $K = [H \quad G]$ for the desired resonance frequency and the desired quality factor can be calculated in MATLAB using the pole placement approach discussed in Section 2.

********* Insert Figure 3 here **************************************

## V. EXPERIMENTS

### V.I. Effect of State Gains on the Probe Response

We investigated the effect of position and velocity gains on the response of the probe. The resonance frequency and the damping characteristics of the system can be set by adjusting the position and velocity gains. In Fig. 4, we show the combined effect of changing velocity and position gains for adjusting the effective $Q$ factor and resonance frequency of the probe simultaneously. Note that the numerical model (blue solid line) shows excellent agreement with the experimental data (red dashed line). As shown in the figure, the resonance frequency of the probe is shifted to the right and its effective $Q$ factor is reduced at the same time to decrease the time constant and hence improve its response time.

********* Insert Figure 4 here **************************************

In Figure 5, the effect of position gain $H$ on the resonance frequency and on the $Q$ factor of the probe are plotted along with the effect of velocity gain $G$ on the $Q$ factor.

********* Insert Figure 5 here **************************************

To further analyze the transient response of the probe, the effect of altering $G$ and $H$ gains on the time constant of the probe, $\tau$, is also investigated using the numerical model. Fig. 6 shows the percent change in time constant of the probe as the gains $G$ and $H$ are altered. In general, decreasing the position gain $H$ (i.e. increasing the resonance frequency) and/or decreasing the velocity gain *G* (i.e. decreasing the *Q* factor) has a positive effect of reducing the time constant, but the influence of resonance frequency on the time constant is less significant at negative values of gain *G*.

********* Insert Figure 6 here *************************************

**V.II. Effect of State Gains on the Scan Profile**

The effect of gain settings on the scan results is investigated using a SIMULINK model and the numerical simulations are verified via real experiments. For this purpose, the transfer function of the probe developed in Section 4 is inserted into our SIMULINK model (see the details of this model in Varol et al.[13]) and then the feedback loops for position and velocity are added. Using this model, we simulate the process of scanning 100 nm steps (width = 1.5 μm) in tapping mode AFM and investigate the effect of state gains on the scan error (see Fig. 7). The scan error is calculated by first integrating the positional error between the profile obtained from the SIMULINK model and the desired profile over a step width and then normalizing the sum by the area under the step[13]. As shown in Fig. 7, the scan error is reduced as the resonance frequency is increased and/or the *Q* factor is decreased (note that both alterations reduce the time constant of the probe and improve its response time).

********* Insert Figure 7 here *************************************

To verify the numerical simulations, we performed scan experiments in tapping mode using our AFM setup. A calibration grating having periodic steps of 100 nm (TGZ02, range: 94 nm-106 nm, Mikromash, USA) is scanned at two different scan speeds ($v_1 = 4$ μm/s and $v_2 = 20$ μm/s) using the experimental setup. In Fig. 8, we compare the scans obtained by a probe having a high $Q$ factor ($K = [0 \quad 0.3]$) with the scans of a probe having a low $Q$ factor and higher resonance frequency ($K = [-6.5 \quad -1.5]$). As shown in Fig. 8a, the response of the probe having low time constant (red dashed line) is better than the one having higher time constant (blue dashed line). The probe responds faster to the same step when its time constant is reduced using the state feedback control. The response of the probe at high scan speeds is slow and causes an inclined profile for the upward and downward steps (see Fig. 8b). As the scanning speed increases the positive effect of the state feedback control decreases, but still exists.

********* Insert Figure 8 here *************************************

Instead of applying state feedback control to the cantilever probe, one can also increase the proportional gain P of the XYZ scanner to make the upward and downward steps sharper during a scan, but this also results in more oscillations and overshoots in the scanner response and frequent sticking of the probe to the surface (see Fig. 9a). In Figure 9b, we compare the scan results obtained at high proportional gain P for the probe having high (blue solid line) and low (red solid line) time constants. As it can be seen, the state feedback control of the probe decreases the overshoots, prevents the probe from sticking to the surface, and hence results in a better scan profile.

********* Insert Figure 9 here *************************************

While reducing the time constant of the probe (by increasing its resonance frequency and/or reducing the $Q$ factor), decreases the scan error, it also increases the tapping forces applied by the probe on the sample. Since measuring these forces experimentally is highly difficult, we investigate the effect of state gains on the tapping forces through numerical simulations. As shown in Fig. 10, the tapping forces increase as the resonance frequency is increased and/or the $Q$ factor is decreased. The tapping force is calculated using the average of maximum indentations of the probe tip into the sample surface after the tapping amplitude reaches steady state.

********* Insert Figure 10 here ***************************************

## VI. DISCUSSION AND CONCLUSION

We presented an approach for adjusting the resonance frequency and the $Q$ factor of an AFM probe using state feedback control. For this purpose, an analog circuit was built to change the effective stiffness and damping of the probe. While changing the damping affects the $Q$ factor of the probe only, a change in the stiffness modifies both the resonance frequency and the $Q$ factor. For this reason, it is important to change the effective stiffness and the damping of the probe simultaneously using a state feedback controller, which enables us to select the poles of the transfer function representing the probe based on the desired values of the $Q$ factor and $\omega_n$. To implement the state feedback control in dynamic AFM, it was necessary to obtain the full state (position and velocity) of the probe as a function of time. Since the LDV can only measure the velocity of the oscillating probe, we built an analog integrator circuit to obtain the position signal from the measured velocity signal through integration. We then calculated the position and velocity gains ($H$ and $G$) that must be used in the feedback loop to achieve the desired values of $\omega_n$ and $Q$ factor, respectively. For this purpose, we first estimated the transfer function of the probe using the frequency sweep data and then applied the pole placement technique as discussed in Section 2. Through this analysis, we observed that the magnitude of the positional gain *H* in our system is significantly higher than that of the velocity gain *G*. It was obvious that this high gain *H* could not be provided by the voltage supply directly. We provided the major part of this gain through the integrator circuit and the rest is delivered by the voltage supply.

We performed scan experiments with probes having high and low time constants by adjusting the feedback gains. The results of these experiments showed that as the time constant of the probe is reduced (and hence the response of the probe is faster), the scan error is reduced and image quality is improved. While the same effect can be achieved by increasing the proportional gain of the XYZ scanner, this also results in more oscillations and overshoots in scanner response and frequent sticking of the probe to the surface. In order to reduce the time constant of the probe, we simultaneously reduced its *Q* factor and increased its resonance frequency. However, the results of our numerical simulations showed that this also increases the magnitude of the tapping forces, which is not desirable.

Instead of using the pole placement approach to set the feedback gains, an optimum controller can also be designed such that the scan error and/or the tapping forces are minimized. Our study shows that there is a trade-off between the scan error and the tapping forces. Attempting to reduce the scan error by reducing the *Q* factor and increasing $\omega_n$ causes the tapping forces to increase. Hence, an optimum controller can be a remedy to this problem. Moreover, as an alternative to building an analog integrator circuit, a state observer could be used to estimate the position signal from the measured velocity signal. For example, Sebastian et al.[16] estimate the velocity from the measured position signal using a state observer. In our setup, a DAQ card with a high sampling rate is required for the realization of a state observer since the operating frequency of our probe is quite high (around 224kHz). Implementation of a state observer will be the subject of our future research.

**Figures**

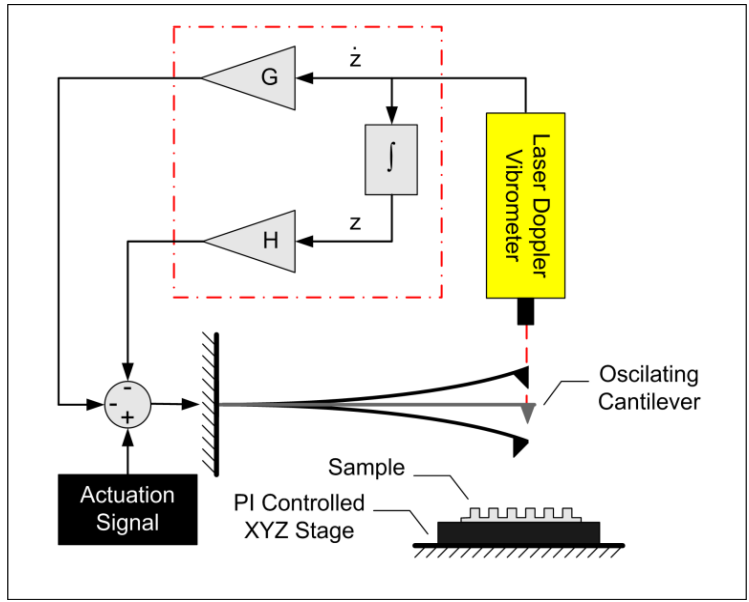

**FIG. 1**. (Color online) State feedback control of a piezo-actuated bimorph AFM probe. The feedback gains *H* and *G* enable us to operate the probe at a desired resonance frequency and *Q* factor. The area marked by the red dashed line is implemented as an analog circuit (see Fig. 2).

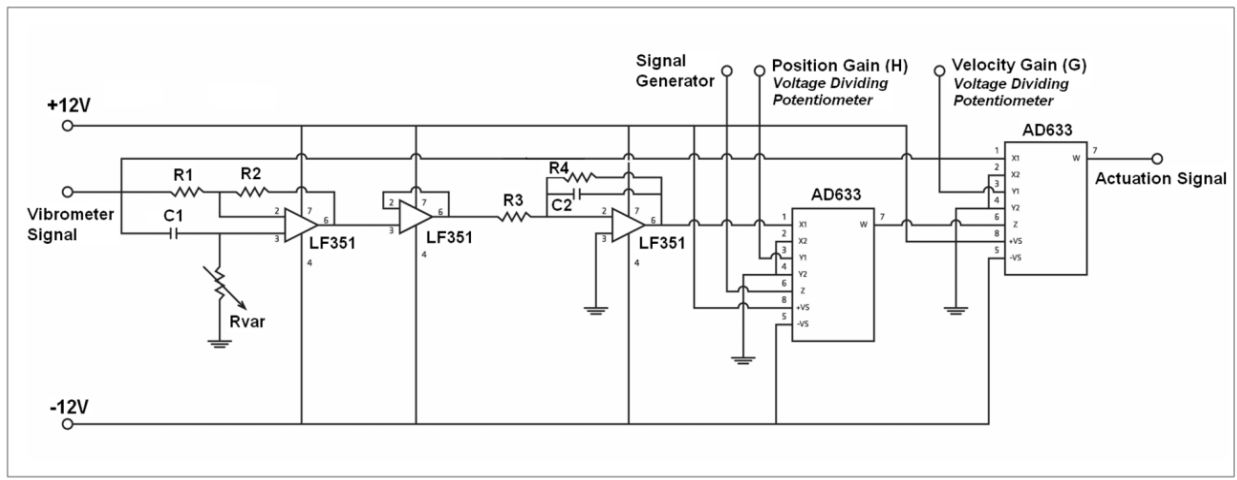

**FIG. 2**. The analog circuit used for the implementation of the state feedback approach.

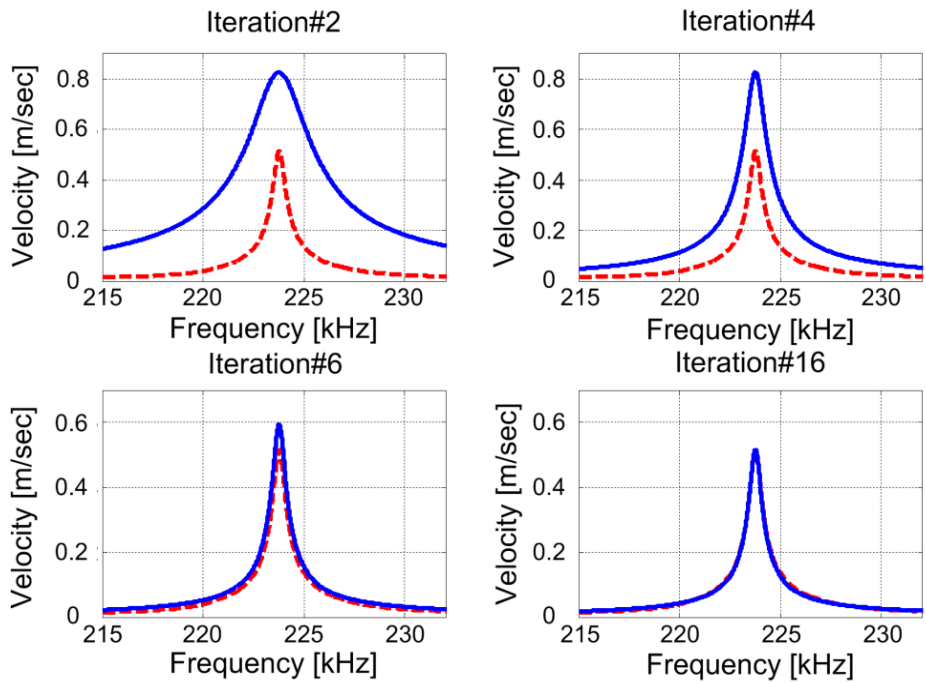

**FIG. 3**. (Color online) Snapshots showing the curve fitting approach used to estimate the parameters of the transfer function of the probe. The red dashed curve represents the experimental data, whereas the blue solid one is the estimated curve.

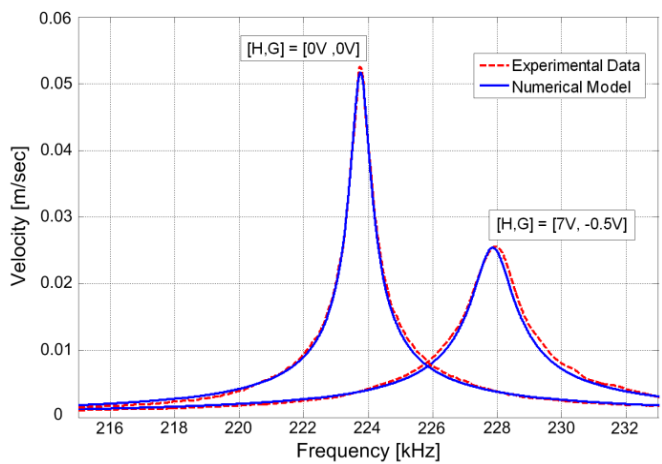

**FIG. 4**. (Color online) The resonance frequency and the $Q$ factor of the probe are set to the desired values by altering the position and velocity gains simultaneously.

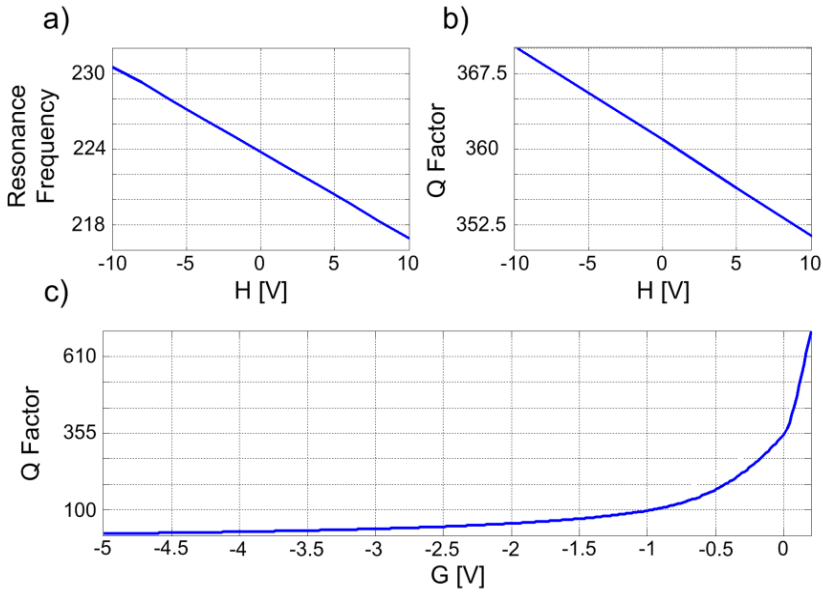

**FIG. 5**. (Color online) The relation (a) between the position gain $H$ and the resonance frequency, (b) between the position gain $H$ and the $Q$ factor, and (c) between the velocity gain $G$ and the $Q$ factor of the probe.

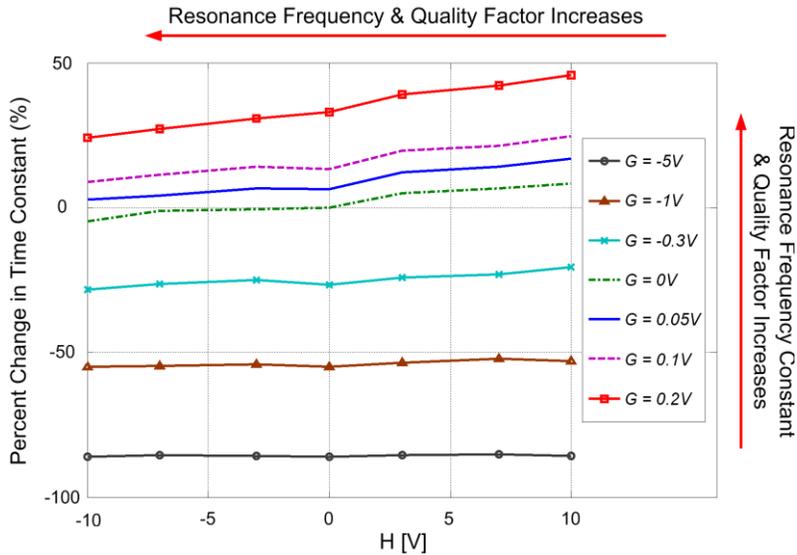

**FIG. 6**. (Color online) The effect of state gains on the time constant of the probe.

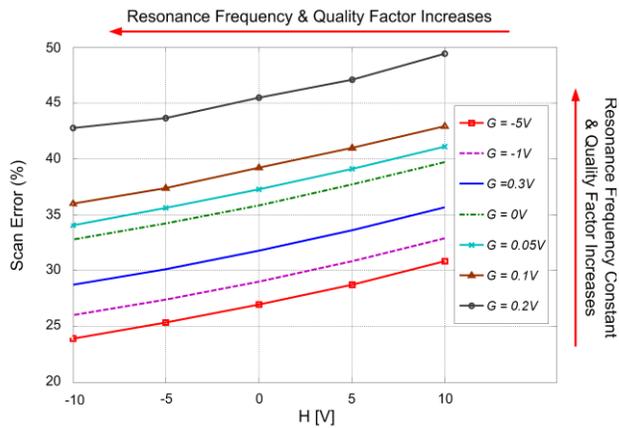

**FIG. 7**. (Color online) The effect of state gains on the scan error is investigated using the SIMULINK model.

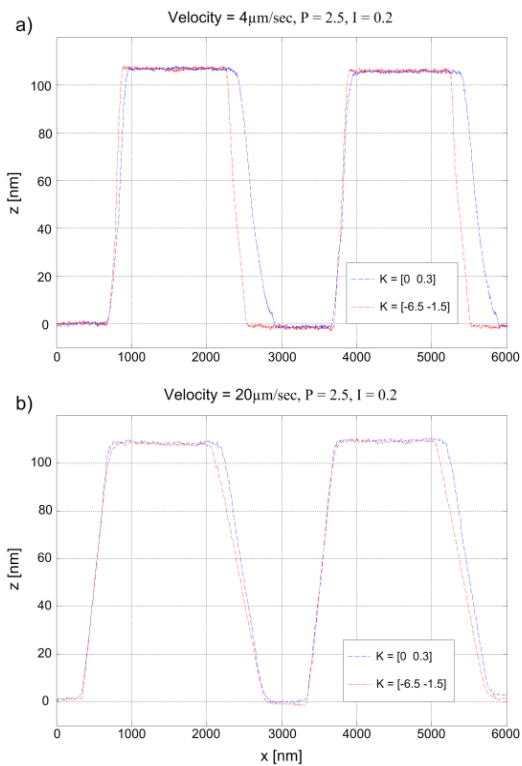

**FIG. 8**. (Color online) The effect of state gains on the scan error is investigated experimentally using a calibration grating having periodic steps. The dashed blue and red lines represent the response of the probe having high and low time constants, respectively. Note that the control parameters for the XYZ scanner are the same in both cases (P = 2.5, I = 0.2).

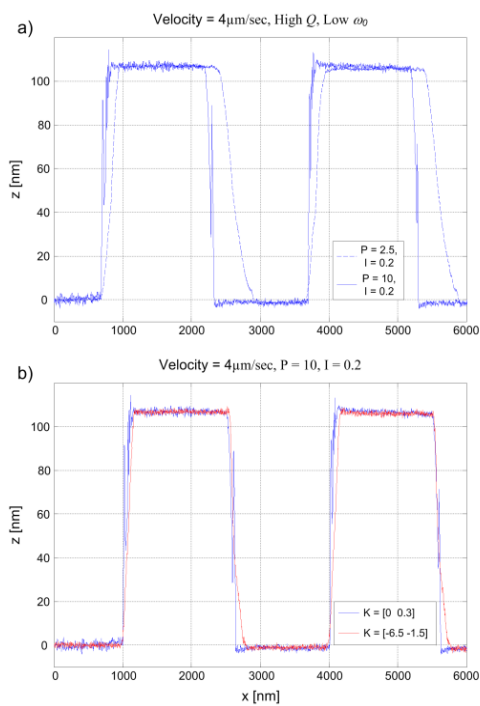

**FIG. 9.** (Color online) (a) The effect of increasing the proportional gain P of the XYZ scanner on the scan results (b) The effect of state feedback control of probe on the scan results at high proportional gain P = 10.0.

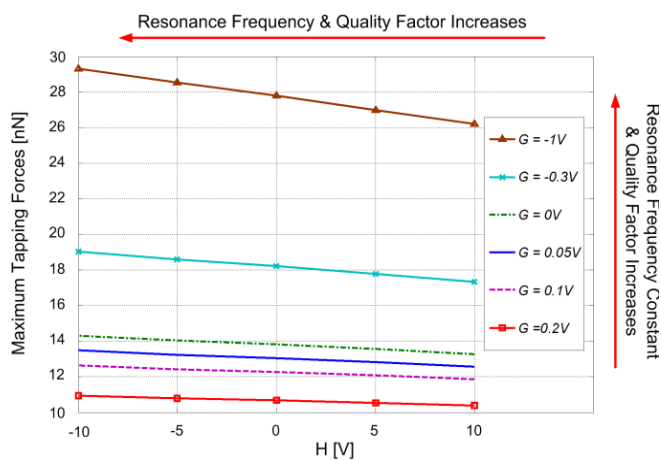

**FIG. 10. (Color online)** The effect of state gains on the tapping forces is investigated using the numerical model.